\documentclass[journal=jacsat,manuscript=article]{achemso}
\usepackage[version=3]{mhchem}

\usepackage[normalem]{ulem}

\usepackage{graphicx}
\usepackage{bm}
\usepackage{color}
\usepackage[normalem]{ulem}
\usepackage[dvipsnames]{xcolor}
\usepackage{cleveref}
\usepackage{amsmath}
\usepackage{textcomp}
\usepackage[utf8]{inputenc}

\newcommand{\be}{\begin{equation}}
\newcommand{\ee}{\end{equation}}
\newcommand{\bea}{\begin{eqnarray}}
\newcommand{\eea}{\end{eqnarray}}

\def\wse2{WSe$_2$}
\def\mose2{MoSe$_2$}
\def\moire{moir\'e }
\def\didv{$dI/dV$}

\usepackage[symbol]{footmisc}

\usepackage{changes}

\author{Sara Shabani}
\affiliation{Department of Physics, Columbia University, New York, NY, USA}
\author{Thomas P. Darlington}
\affiliation{Department of
Mechanical Engineering, Columbia University, New York, NY, USA}
\author{Colin Gordon}
\affiliation{Department of Physics and Astronomy, Stony Brook University, Stony Brook, NY, USA}
\author{Wenjing Wu}
\affiliation{Department of Chemistry, Columbia University, New York, NY, USA}
\author{Emanuil Yanev}
\affiliation{Department of
Mechanical Engineering, Columbia University, New York, NY, USA}
\author{James Hone}
\affiliation{Department of Mechanical Engineering, Columbia University, New York, NY, USA}
\author{Xiaoyang Zhu}
\affiliation{Department of Chemistry, Columbia University, New York, NY, USA}
\author{Cyrus E. Dreyer}
\email{cyrus.dreyer@stonybrook.edu}
\affiliation{Center for Computational Quantum Physics, Flatiron Institute, New York, NY, USA}
\affiliation{Department of Physics and Astronomy, Stony Brook University, Stony Brook, NY, USA}
\author{P. James Schuck}
\email{pjs2191@columbia.edu}
\affiliation{Department of
Mechanical Engineering, Columbia University, New York, NY, USA}
\author{\hspace{0.11 cm}Abhay N. Pasupathy}
\email{apn2108@columbia.edu}
\affiliation{Department of Physics, Columbia University, New York, NY, USA}

\title{Ultra Localized Optoelectronic Properties of Nanobubbles in 2D Semiconductors}

\begin{document}

\maketitle
\begin{abstract}
The optical properties of transition metal dichalcogenides have previously been modified at the nanoscale by using mechanical and electrical nanostructuring. However, a clear experimental picture relating the local electronic structure with emission properties in such structures has so far been lacking. Here, we use a combination of scanning tunneling microscopy (STM) and near-field photoluminescence (nano-PL) to probe the electronic and optical properties of single nano-bubbles in bilayer heterostructures of \wse2 on \mose2.  We show from tunneling spectroscopy that there are electronic states deeply localized in the gap at the edge of such bubbles, which are independent of the presence of chemical defects in the layers. We also show a significant change in the local bandgap on the bubble, with a continuous evolution to the edge of the bubble over a length scale of $\sim$20 nm. Nano-PL measurements observe a continuous redshift of the interlayer exciton on entering the bubble, in agreement with the band to band transitions measured by STM. We use self-consistent Schrödinger-Poisson (SP) simulations to capture the essence of  the experimental results, and find that strong doping in the bubble region is a key ingredient to achieving the observed localized states, together with mechanical strain.
\end{abstract}
\section{Introduction}
Two-dimensional (2d) transition metal dichalcogenides (TMDs) have been investigated as a possible route toward producing well-controlled optical emitters, critical components in quantum technologies and photonics \cite{gabel2021imaging,dhakal2017local,branny2017deterministic,cho2021highly,turunen2022quantum,XYZ:2020:excitonic,chowdhury2021anomalous,koo2021tip,lee2021inducing,luo2020exciton,parto2021defect,basov2022nano}. The bandgap of monolayer TMDs is uniquely sensitive to doping\cite{xiaodong2019visualizing,kim2019electrical,liu2019direct,yao2017optically,chernikov2015population}, dielectric environment\cite{Heinz2014measurement} and mechanical strain\cite{li2015optoelectronic,Drew:2020,koo2021tip,parto2021defect,trainer2019effects}, thereby providing an ideal platform to create quantum confinement  by local mechanical and electrical fields. Indeed, previous research has investigated moire patterns \cite{kim2021excitons,xiaodong2021moire,shabani2021deep,XYZ:2019:1d,crommie2021correlated}, wrinkles \cite{cho2021highly,dhakal2017local,koo2021tip}, bubbles \cite{gabel2021imaging,darlington2020imaging} and lithographically fabricated dielectric arrays as effective methods \cite{carmesin2019quantum} to create localized optical emitters. In spite of these notable achievements, a basic band diagram picture across such localized structures remains elusive. For example, recent theoretical calculations\cite{chirolli2019strain,carmesin2019quantum,morrow2021trapping} have predicted large bandgap changes at the edges of wrinkles and nanobubbles. However, such predictions have not been experimentally verified, and it remains unclear what the major contributors are to achieving localized optical emitters in TMD materials.

In order to address these questions, high spatial resolution spectroscopic measurements across local perturbations in TMDs are necessary. Our approach is to use the local imaging and spectroscopic capabilities of  scanning tunneling microscopy(STM) to measure the local band diagram across nano-bubbles, and directly correlate these measurements with near-field optical measurements with ~10 nm spatial resolution.

\section{Results and Discussion}
Our STM measurements are conducted on nearly aligned (61.7 $^{\circ}$) \wse2/\mose2 heterobilayers that were stacked on Graphite/hBN as a conducting electrode for STM measurements (see Fig.~1a for a schematic). The relative orientation between two TMD monolayers was determined by second harmonic generation (SHG) (see supporting information figure S1 and accompanying discussion). The  details of sample fabrication are presented in methods. During the stacking process, nanobubbles are commonly formed between layers. Figure 1b shows a large scale atomic force microscopic (AFM ) image of such nanobubbles. The nanobubbles range in size from 20 nm to 400 nm and in height from 2 nm to 50 nm. STM topographic measurements on one such nanobubble are shown in Fig. 1c.  Figure 1d  shows height profile of the nanobubble boundary corresponding to the dashed arrow in Fig.~1c. In order to obtain the STM topography, the tunneling current remains constant at a particular voltage bias, through a feedback loop; however, the tunneling current depends on the integrated local density of states between the sample Fermi level and the bias voltage. This dependency of current on electronic structure affects the topographic data. Thus, both electronic structure and actual topographic height influence the apparent topographic height in STM and the apparent heights are not directly translatable into real vertical displacements. This consideration is not relevant to AFM measurements; however, AFM measurements are dictated by tip-sample force interactions which can strongly influence the apparent height in AFM topographic scans.  However, it is clearly evident from the line profile across the boundary shown in Fig. 1d that the edge of the bubble features a sharp step. Such a sharp step indicates that there is localized strain at the edge of the bubble. Finally, a \moire pattern is clearly visible both on and off the bubble. Its presence in all regions shows that the \wse2 and \mose2 layers are in good contact throughout, and the bubble is under the heterobilayer. As a result of the device fabrication process under ambient conditions, compounds in air such as water molecules often are trapped between layers. While the chemical identity of the material in the bubble is unknown, our data below, which shows a highly doped region on the nanobubble, indicate the presence of polarized (or ionized) molecules trapped in the nanobubble during the fabrication. Such polarized molecules induce doping and electric field to the layers above and modify the electronic structure.

To characterize the electronic properties in the vicinity of the nanobubble, we perform STM \didv{} spectroscopy measurements at various points on and off the bubble (see Fig. 2a). Shown in Fig.~2b and c are a sequence of such measurements taken across the edge of the bubble, color coded according to the markers on Fig. 2a. Figure 2b shows the typical spectra well inside (blue) and outside (red) the nanobubble area. It is seen that the conduction band (CB) edge shows a substantial shift from the outer edge of the bubble into the interior while VB is nearly unchanged. This indicates a substantial reduction in the bandgap on the bubble as well a large electron doping.

Shown in Fig.~2c are a sequence of spectra taken in the transition region across the bubble edge, with a focus on the CB edge where the most dramatic changes are observed. It is seen that at the edge of the bubble (red curve), electronic states exist that are deeply bound in the semiconductor gap. As we transition from outside to inside the bubble (green curve), these states move towards the CB edge, while at the same time the CB edge moves towards the Fermi level. The entire evolution is shown as a heat map of \didv{} in Fig.~2d as a function of position along the arrow overlaid on Fig.~2a. The  dashed lines overlaid on Fig.~2d indicate the evolution of the localized states and the conduction band edge upon moving from the outside to the inside the bubble.  

In order to confirm that the observations above are representative of the entire bubble and not specific to a particular region, we performed spectroscopic imaging experiments across the bubble interface at various energies, a subset of which are shown in Fig.~2e-g. At the valence band edge (Fig.~2e), the map does not distinguish between the interior and exterior of the bubble, showing the uniformity of the valence band edge across the bubble. Figure 2f and 2g, taken within the semiconducting gap, show the presence of the localized states clearly, and their evolution towards the interior of the bubble. At large positive energies (not shown), these localized states merge into the conduction band, and a clear contrast is seen between the interior and exterior of the bubble. These maps confirm the basic picture provided by point spectra -- the bubble region is characterized by localized states at its edge and a shifted conduction band edge throughout. We present larger maps across the entire bubble in the supporting information, to rule out moir\'e effects on the observed phenomena. Finally, to rule out possible damage to the bubble during spectroscopic maps, we measure the topographic heights before and after spectroscopy is performed.

We next proceed to compare our measurements of the single particle spectroscopic properties of individual bubbles with their optical emission. To do this, we employed hyperspectral nanoscale photoluminescence measurements to map the changes in exciton energy across the bubble. We have previously applied this technique to the imaging of localized exciton states in transition metal dichalcogenide monolayers \cite{darlington2020imaging}, however the spatial resolution was limited due to the finite radius of curvature of the probe. To push the nano-optical resolution to scales on order the localized state, we prepared hetero-nanobubbles on template stripped gold (TS-Au), which allows for optical resolutions on order of the gap formed between the nano-optic probe and the substrate \cite{Khoury2020acs}. In addition, the quenching of photoluminence of the monolayer and heterobilayer regions in contact with the substrate greatly reduces background, improving nanobubble's PL contrast \cite{tyurnina2019strained}.

Figure 3a shows an AFM image of a nanobubble \wse2/\mose2 on TS-Au, that shows localized interlayer exciton emission (Fig. 3b) when compared with the flat heterostructure. The greater strength of the interlayer exciton emission on the nanobubble edge can be attributed to two factors. First, there is an enhancement from the strain of the nano-bubble itself, which shifts the absorption of both constituent layers to the red, allowing for greater absorption of the excitation photons. Second, recent work has shown that like their intra-layer cousins, the interlayer excitons transition dipole is primarily in the plane of the 2D layers\cite{sigl2022optical}. As the nano-optical probe moves across the edge of the nanobubble, more of the transition dipole is aligned with the polarization of the nano-optical field, allowing for significantly greater in- and out-coupling for optical fields. This alignment we believe is critical for near-field observation of inter-layer exciton luminescence, which has substantially weaker oscillator strength. We also expect that strain-localized PL signal is enhanced by funneling effects, where excitons are preferentially shuttled towards the lower energy states\cite{su2022dark,harats2020dynamics,lee2022drift}

Figure 3c shows point spectra at various locations across the nanobubble edges, identified by the colored markers in Fig.~3a and 3d. Black arrows show visual identification of the localized interlayer exciton peak. Clear red shifting of the emission is observed as one moves from the outer edge of the bubble into the interior. To visualize this more clearly, in Fig.~3c we plot a hyperspectral linescan along the vector defined by colored points in Fig.~3a. The general trend is towards redder emission, shifting ~ 200 meV over 15 nm into the bubble, after which the emission quickly falls to zero in the nanobubble interior. In Figs.~3e - 3g we show this same trend with spatial maps with energy bins of 1.40, 1.30, and  1.25 eV, respectively. The size of the localized interlayer exciton emission “ring” clearly shrinks for the redder energy bins. 

The shifting of the localized interlayer exciton emission energy reflects the energic shift of the \mose2 conduction band observed in the STS. Indeed, comparing the STS and nano-PL hyperspectral linescans, the roll off of the interlayer emission energy and the conduction band shows remarkable similarities at the nanobubble, but the nano-PL shows a sudden drop-off in emission intensity soon after the edge, which is not seen in STS. This drop-off however is expected due to the silicon detector used for the nano-PL spectroscopy (see methods for details), which rapidly lose quantum-efficiency at energies lower than $\sim$1.3 eV. Similarly, we do not observe emission from the deeply localized defect states seen in Fig.~2 given that this would correspond to emission energies $\sim$0.7 eV, which is far below the cutoff our detector. While quantitative comparison between the STM and nano-PL data is not possible due to the sample differences, evaluation of the STS-derived band gap and exciton PL energy provides information on the binding energy for excitons associated with the conduction band near the bubble edge. Our results are consistent with a conduction band interlayer exciton binding energy with magnitude of a couple hundred meV within the bubble\cite{kamban2020interlayer}.

To understand the contributions of doping and strain
to the observed spectroscopic features in experiment, we
performed Schr\"odinger-Poisson (SP) simulations
using the approach and code of Ref. ~\cite{bussy2017strain}. The
lengthscale associated with the entire bubble region is
too large to model directly, but we will show that the
pertinent experimental features can be captured with a
significantly simplified model. Our focus on the
conduction band of the heterostructure, which is
derived from the \mose2 layer. Thus, we take band
parameters for \mose2 from Ref. ~\cite{kormanyos2015k} and
electromechanical parameters from Ref. ~\cite{duerloo2012intrinsic}. Since the
important physics is localized to the bubble edge, we
approximate the bubble region by a 2D inclusion in monolayer
\mose2 (large enough to isolate the two interfaces
required by periodic boundary conditions), with a
downward shift of the conduction band in the bubble
region, as well as a background doping, that both
increase with strain \cite{Listrain}. The strain is assumed to be
slightly larger at the edge of the bubble region. The
DOS is calculated assuming a 2D parabolic dispersion
in the direction parallel to the inclusion above the
conduction-band minimum and below the valence band maximum.

Figure 4a show the band diagram obtained by self-consistent solution of the SP equations, focusing on the CB. By construction, the Fermi level is at the conduction band within the bubble region due to the background doping. The dips in the electrostatic potential of the conduction-band at the edges of the ``bubble'' have two origins. The first is the band bending induced by the interface between the doped and undoped region. The second is that we assume a higher strain at the interface, which increases the effect of this band bending. We found that adding a piezoelectric polarization charge at the interface (not shown) did not change the qualitative features, other than introducing an asymmetry in the band-bending on the opposite sides of the bubble region. 

The red dot-dashed lines in Fig.~4a are the energies of the bound states in the bubble region, and the green dotted lines are their squared wavefunctions (shifted so that the zero-density level is at the energy for clarity). We can see that the band bending at the interface region results in a bound state, which can also be clearly seen in Fig.~4b as spikes in the electron density at the edge of the bubble regions. In Fig.~4c we plot the DOS of the total system, including the sum of 1D states in the bubble region, and the 2D DOS in the bulk, at different x points near the interface of the bubble (see the inset, denoted by curve color). Superimposed on the step-like increase from the 2D DOS at the conduction and valence band edges in the bubble, we can see a small enhancement in the DOS near the interface, where the localized state resides. As we move toward the center of the bubble region, this enhancement weakens, and ultimately disappears.
We focused our model on the \mose2 layer of the
bilayer structure since the \mose2 is the main
contributor to the conduction band. The size of the
bubble is much larger than the crystal cell. If you were
to zoom in at the atomic level the edge of the bubble
can be approximated by its tangent on the surface of
the \mose2 layer.
Hence using the deformation gradients as a baseline
for the energy of the different strained portions we
can accurately model the band edge with the one-
dimensional model.
\section{Conclusion}
Our theoretical results, though the result of a simplified model, accurately captures the essential features of the experimental observations. They indicate that sharp lateral junctions in doping are an excellent way to engineer localized states in 2D TMD semiconductors. Such localized states do not depend on the presence of specific chemical defects or specialized strain fields. Recently, several 2D materials interfaces have shown the ability to realize large charge transfer at the interface \cite{rucl3,balgley2022ultra}. Our results indicate that such interfacial engineering together with nanostructuring can be employed to create optical emitters with arbitrarily desired shapes at the nanoscale. Our work provides a clear route to achieving this in the future.

\bibliography{biba.bib}

\providecommand{\latin}[1]{#1}
\makeatletter
\providecommand{\doi}
  {\begingroup\let\do\@makeother\dospecials
  \catcode`\{=1 \catcode`\}=2 \doi@aux}
\providecommand{\doi@aux}[1]{\endgroup\texttt{#1}}
\makeatother
\providecommand*\mcitethebibliography{\thebibliography}
\csname @ifundefined\endcsname{endmcitethebibliography}
  {\let\endmcitethebibliography\endthebibliography}{}
\begin{mcitethebibliography}{44}
\providecommand*\natexlab[1]{#1}
\providecommand*\mciteSetBstSublistMode[1]{}
\providecommand*\mciteSetBstMaxWidthForm[2]{}
\providecommand*\mciteBstWouldAddEndPuncttrue
  {\def\EndOfBibitem{\unskip.}}
\providecommand*\mciteBstWouldAddEndPunctfalse
  {\let\EndOfBibitem\relax}
\providecommand*\mciteSetBstMidEndSepPunct[3]{}
\providecommand*\mciteSetBstSublistLabelBeginEnd[3]{}
\providecommand*\EndOfBibitem{}
\mciteSetBstSublistMode{f}
\mciteSetBstMaxWidthForm{subitem}{(\alph{mcitesubitemcount})}
\mciteSetBstSublistLabelBeginEnd
  {\mcitemaxwidthsubitemform\space}
  {\relax}
  {\relax}

\bibitem[Gabel \latin{et~al.}(2021)Gabel, El-Khoury, and Gu]{gabel2021imaging}
Gabel,~M.; El-Khoury,~P.~Z.; Gu,~Y. Imaging Charged Exciton Localization in van
  der Waals WSe2/MoSe2 Heterobilayers. \emph{The Journal of Physical Chemistry
  Letters} \textbf{2021}, \emph{12}, 10589--10594\relax
\mciteBstWouldAddEndPuncttrue
\mciteSetBstMidEndSepPunct{\mcitedefaultmidpunct}
{\mcitedefaultendpunct}{\mcitedefaultseppunct}\relax
\EndOfBibitem
\bibitem[Dhakal \latin{et~al.}(2017)Dhakal, Roy, Jang, Chen, Yun, Kim, Lee,
  Kim, and Ahn]{dhakal2017local}
Dhakal,~K.~P.; Roy,~S.; Jang,~H.; Chen,~X.; Yun,~W.~S.; Kim,~H.; Lee,~J.;
  Kim,~J.; Ahn,~J.-H. Local strain induced band gap modulation and
  photoluminescence enhancement of multilayer transition metal dichalcogenides.
  \emph{Chemistry of Materials} \textbf{2017}, \emph{29}, 5124--5133\relax
\mciteBstWouldAddEndPuncttrue
\mciteSetBstMidEndSepPunct{\mcitedefaultmidpunct}
{\mcitedefaultendpunct}{\mcitedefaultseppunct}\relax
\EndOfBibitem
\bibitem[Branny \latin{et~al.}(2017)Branny, Kumar, Proux, and
  Gerardot]{branny2017deterministic}
Branny,~A.; Kumar,~S.; Proux,~R.; Gerardot,~B.~D. Deterministic strain-induced
  arrays of quantum emitters in a two-dimensional semiconductor. \emph{Nature
  communications} \textbf{2017}, \emph{8}, 1--7\relax
\mciteBstWouldAddEndPuncttrue
\mciteSetBstMidEndSepPunct{\mcitedefaultmidpunct}
{\mcitedefaultendpunct}{\mcitedefaultseppunct}\relax
\EndOfBibitem
\bibitem[Cho \latin{et~al.}(2021)Cho, Wong, Taqieddin, Biswas, Aluru, Nam, and
  Atwater]{cho2021highly}
Cho,~C.; Wong,~J.; Taqieddin,~A.; Biswas,~S.; Aluru,~N.~R.; Nam,~S.;
  Atwater,~H.~A. Highly strain-tunable interlayer excitons in MoS2/WSe2
  heterobilayers. \emph{Nano letters} \textbf{2021}, \emph{21},
  3956--3964\relax
\mciteBstWouldAddEndPuncttrue
\mciteSetBstMidEndSepPunct{\mcitedefaultmidpunct}
{\mcitedefaultendpunct}{\mcitedefaultseppunct}\relax
\EndOfBibitem
\bibitem[Turunen \latin{et~al.}(2022)Turunen, Brotons-Gisbert, Dai, Wang,
  Scerri, Bonato, J{\"o}ns, Sun, and Gerardot]{turunen2022quantum}
Turunen,~M.; Brotons-Gisbert,~M.; Dai,~Y.; Wang,~Y.; Scerri,~E.; Bonato,~C.;
  J{\"o}ns,~K.~D.; Sun,~Z.; Gerardot,~B.~D. Quantum photonics with layered 2D
  materials. \emph{Nature Reviews Physics} \textbf{2022}, 1--18\relax
\mciteBstWouldAddEndPuncttrue
\mciteSetBstMidEndSepPunct{\mcitedefaultmidpunct}
{\mcitedefaultendpunct}{\mcitedefaultseppunct}\relax
\EndOfBibitem
\bibitem[Wang \latin{et~al.}(2021)Wang, Shi, Shih, Zhou, Wu, Bai, Rhodes,
  Barmak, Hone, Dean, and Zhu]{XYZ:2020:excitonic}
Wang,~J.; Shi,~Q.; Shih,~E.-M.; Zhou,~L.; Wu,~W.; Bai,~Y.; Rhodes,~D.;
  Barmak,~K.; Hone,~J.; Dean,~C.~R.; Zhu,~X.-Y. Diffusivity Reveals Three
  Distinct Phases of Interlayer Excitons in
  ${\mathrm{MoSe}}_{2}/{\mathrm{WSe}}_{2}$ Heterobilayers. \emph{Phys. Rev.
  Lett.} \textbf{2021}, \emph{126}, 106804\relax
\mciteBstWouldAddEndPuncttrue
\mciteSetBstMidEndSepPunct{\mcitedefaultmidpunct}
{\mcitedefaultendpunct}{\mcitedefaultseppunct}\relax
\EndOfBibitem
\bibitem[Chowdhury \latin{et~al.}(2021)Chowdhury, Jo, Anantharaman,
  Brintlinger, Jariwala, and Kempa]{chowdhury2021anomalous}
Chowdhury,~T.; Jo,~K.; Anantharaman,~S.~B.; Brintlinger,~T.~H.; Jariwala,~D.;
  Kempa,~T.~J. Anomalous Room-Temperature Photoluminescence from Nanostrained
  MoSe2 Monolayers. 2021\relax
\mciteBstWouldAddEndPuncttrue
\mciteSetBstMidEndSepPunct{\mcitedefaultmidpunct}
{\mcitedefaultendpunct}{\mcitedefaultseppunct}\relax
\EndOfBibitem
\bibitem[Koo \latin{et~al.}(2021)Koo, Kim, Choi, Lee, Choi, Lee, Kang, Lee,
  Kim, Lee, \latin{et~al.} others]{koo2021tip}
Koo,~Y.; Kim,~Y.; Choi,~S.~H.; Lee,~H.; Choi,~J.; Lee,~D.~Y.; Kang,~M.;
  Lee,~H.~S.; Kim,~K.~K.; Lee,~G., \latin{et~al.}  Tip-Induced Nano-Engineering
  of Strain, Bandgap, and Exciton Funneling in 2D Semiconductors.
  \emph{Advanced Materials} \textbf{2021}, \emph{33}, 2008234\relax
\mciteBstWouldAddEndPuncttrue
\mciteSetBstMidEndSepPunct{\mcitedefaultmidpunct}
{\mcitedefaultendpunct}{\mcitedefaultseppunct}\relax
\EndOfBibitem
\bibitem[Lee \latin{et~al.}(2021)Lee, Kim, Park, Kang, Choi, Jeong, Mun, Kim,
  Park, Raschke, \latin{et~al.} others]{lee2021inducing}
Lee,~H.; Kim,~I.; Park,~C.; Kang,~M.; Choi,~J.; Jeong,~K.-Y.; Mun,~J.; Kim,~Y.;
  Park,~J.; Raschke,~M.~B., \latin{et~al.}  Inducing and Probing Localized
  Excitons in Atomically Thin Semiconductors via Tip-Enhanced
  Cavity-Spectroscopy. \emph{Advanced Functional Materials} \textbf{2021},
  \emph{31}, 2102893\relax
\mciteBstWouldAddEndPuncttrue
\mciteSetBstMidEndSepPunct{\mcitedefaultmidpunct}
{\mcitedefaultendpunct}{\mcitedefaultseppunct}\relax
\EndOfBibitem
\bibitem[Luo \latin{et~al.}(2020)Luo, Liu, Kim, Hone, and
  Strauf]{luo2020exciton}
Luo,~Y.; Liu,~N.; Kim,~B.; Hone,~J.; Strauf,~S. Exciton dipole orientation of
  strain-induced quantum emitters in WSe2. \emph{Nano letters} \textbf{2020},
  \emph{20}, 5119--5126\relax
\mciteBstWouldAddEndPuncttrue
\mciteSetBstMidEndSepPunct{\mcitedefaultmidpunct}
{\mcitedefaultendpunct}{\mcitedefaultseppunct}\relax
\EndOfBibitem
\bibitem[Parto \latin{et~al.}(2021)Parto, Azzam, Banerjee, and
  Moody]{parto2021defect}
Parto,~K.; Azzam,~S.~I.; Banerjee,~K.; Moody,~G. Defect and strain engineering
  of monolayer WSe2 enables site-controlled single-photon emission up to 150 K.
  \emph{Nature communications} \textbf{2021}, \emph{12}, 1--8\relax
\mciteBstWouldAddEndPuncttrue
\mciteSetBstMidEndSepPunct{\mcitedefaultmidpunct}
{\mcitedefaultendpunct}{\mcitedefaultseppunct}\relax
\EndOfBibitem
\bibitem[Zhang \latin{et~al.}(2022)Zhang, Li, Chen, Ruta, Shao, Sternbach,
  McLeod, Sun, Xiong, Moore, \latin{et~al.} others]{basov2022nano}
Zhang,~S.; Li,~B.; Chen,~X.; Ruta,~F.~L.; Shao,~Y.; Sternbach,~A.~J.;
  McLeod,~A.; Sun,~Z.; Xiong,~L.; Moore,~S., \latin{et~al.}  Nano-spectroscopy
  of excitons in atomically thin transition metal dichalcogenides. \emph{Nature
  Communications} \textbf{2022}, \emph{13}, 1--8\relax
\mciteBstWouldAddEndPuncttrue
\mciteSetBstMidEndSepPunct{\mcitedefaultmidpunct}
{\mcitedefaultendpunct}{\mcitedefaultseppunct}\relax
\EndOfBibitem
\bibitem[Nguyen \latin{et~al.}(2019)Nguyen, Teutsch, Wilson, Kahn, Xia, Graham,
  Kandyba, Giampietri, Barinov, Constantinescu, \latin{et~al.}
  others]{xiaodong2019visualizing}
Nguyen,~P.~V.; Teutsch,~N.~C.; Wilson,~N.~P.; Kahn,~J.; Xia,~X.; Graham,~A.~J.;
  Kandyba,~V.; Giampietri,~A.; Barinov,~A.; Constantinescu,~G.~C.,
  \latin{et~al.}  Visualizing electrostatic gating effects in two-dimensional
  heterostructures. \emph{Nature} \textbf{2019}, \emph{572}, 220--223\relax
\mciteBstWouldAddEndPuncttrue
\mciteSetBstMidEndSepPunct{\mcitedefaultmidpunct}
{\mcitedefaultendpunct}{\mcitedefaultseppunct}\relax
\EndOfBibitem
\bibitem[Jauregui \latin{et~al.}(2019)Jauregui, Joe, Pistunova, Wild, High,
  Zhou, Scuri, De~Greve, Sushko, Yu, \latin{et~al.} others]{kim2019electrical}
Jauregui,~L.~A.; Joe,~A.~Y.; Pistunova,~K.; Wild,~D.~S.; High,~A.~A.; Zhou,~Y.;
  Scuri,~G.; De~Greve,~K.; Sushko,~A.; Yu,~C.-H., \latin{et~al.}  Electrical
  control of interlayer exciton dynamics in atomically thin heterostructures.
  \emph{Science} \textbf{2019}, \emph{366}, 870--875\relax
\mciteBstWouldAddEndPuncttrue
\mciteSetBstMidEndSepPunct{\mcitedefaultmidpunct}
{\mcitedefaultendpunct}{\mcitedefaultseppunct}\relax
\EndOfBibitem
\bibitem[Liu \latin{et~al.}(2019)Liu, Ziffer, Hansen, Wang, and
  Zhu]{liu2019direct}
Liu,~F.; Ziffer,~M.~E.; Hansen,~K.~R.; Wang,~J.; Zhu,~X. Direct determination
  of band-gap renormalization in the photoexcited monolayer MoS 2.
  \emph{Physical review letters} \textbf{2019}, \emph{122}, 246803\relax
\mciteBstWouldAddEndPuncttrue
\mciteSetBstMidEndSepPunct{\mcitedefaultmidpunct}
{\mcitedefaultendpunct}{\mcitedefaultseppunct}\relax
\EndOfBibitem
\bibitem[Yao \latin{et~al.}(2017)Yao, Yan, Kahn, Suslu, Liang, Barnard, Tongay,
  Zettl, Borys, and Schuck]{yao2017optically}
Yao,~K.; Yan,~A.; Kahn,~S.; Suslu,~A.; Liang,~Y.; Barnard,~E.~S.; Tongay,~S.;
  Zettl,~A.; Borys,~N.~J.; Schuck,~P.~J. Optically discriminating
  carrier-induced quasiparticle band gap and exciton energy renormalization in
  monolayer MoS 2. \emph{Physical Review Letters} \textbf{2017}, \emph{119},
  087401\relax
\mciteBstWouldAddEndPuncttrue
\mciteSetBstMidEndSepPunct{\mcitedefaultmidpunct}
{\mcitedefaultendpunct}{\mcitedefaultseppunct}\relax
\EndOfBibitem
\bibitem[Chernikov \latin{et~al.}(2015)Chernikov, Ruppert, Hill, Rigosi, and
  Heinz]{chernikov2015population}
Chernikov,~A.; Ruppert,~C.; Hill,~H.~M.; Rigosi,~A.~F.; Heinz,~T.~F. Population
  inversion and giant bandgap renormalization in atomically thin WS 2 layers.
  \emph{Nature Photonics} \textbf{2015}, \emph{9}, 466--470\relax
\mciteBstWouldAddEndPuncttrue
\mciteSetBstMidEndSepPunct{\mcitedefaultmidpunct}
{\mcitedefaultendpunct}{\mcitedefaultseppunct}\relax
\EndOfBibitem
\bibitem[Li \latin{et~al.}(2014)Li, Chernikov, Zhang, Rigosi, Hill, Van
  Der~Zande, Chenet, Shih, Hone, and Heinz]{Heinz2014measurement}
Li,~Y.; Chernikov,~A.; Zhang,~X.; Rigosi,~A.; Hill,~H.~M.; Van
  Der~Zande,~A.~M.; Chenet,~D.~A.; Shih,~E.-M.; Hone,~J.; Heinz,~T.~F.
  Measurement of the optical dielectric function of monolayer transition-metal
  dichalcogenides: MoS 2, Mo S e 2, WS 2, and WS e 2. \emph{Physical Review B}
  \textbf{2014}, \emph{90}, 205422\relax
\mciteBstWouldAddEndPuncttrue
\mciteSetBstMidEndSepPunct{\mcitedefaultmidpunct}
{\mcitedefaultendpunct}{\mcitedefaultseppunct}\relax
\EndOfBibitem
\bibitem[Li \latin{et~al.}(2015)Li, Contryman, Qian, Ardakani, Gong, Wang,
  Weisse, Lee, Zhao, Ajayan, \latin{et~al.} others]{li2015optoelectronic}
Li,~H.; Contryman,~A.~W.; Qian,~X.; Ardakani,~S.~M.; Gong,~Y.; Wang,~X.;
  Weisse,~J.~M.; Lee,~C.~H.; Zhao,~J.; Ajayan,~P.~M., \latin{et~al.}
  Optoelectronic crystal of artificial atoms in strain-textured molybdenum
  disulphide. \emph{Nature communications} \textbf{2015}, \emph{6}, 1--7\relax
\mciteBstWouldAddEndPuncttrue
\mciteSetBstMidEndSepPunct{\mcitedefaultmidpunct}
{\mcitedefaultendpunct}{\mcitedefaultseppunct}\relax
\EndOfBibitem
\bibitem[Edelberg \latin{et~al.}(2020)Edelberg, Kumar, Shenoy, Ochoa, and
  Pasupathy]{Drew:2020}
Edelberg,~D.; Kumar,~H.; Shenoy,~V.; Ochoa,~H.; Pasupathy,~A.~N. Tunable strain
  soliton networks confine electrons in van der Waals materials. \emph{Nature
  Physics} \textbf{2020}, 1--6\relax
\mciteBstWouldAddEndPuncttrue
\mciteSetBstMidEndSepPunct{\mcitedefaultmidpunct}
{\mcitedefaultendpunct}{\mcitedefaultseppunct}\relax
\EndOfBibitem
\bibitem[Trainer \latin{et~al.}(2019)Trainer, Zhang, Bobba, Xi, Hla, and
  Iavarone]{trainer2019effects}
Trainer,~D.~J.; Zhang,~Y.; Bobba,~F.; Xi,~X.; Hla,~S.-W.; Iavarone,~M. The
  effects of atomic-scale strain relaxation on the electronic properties of
  monolayer MoS2. \emph{ACS nano} \textbf{2019}, \emph{13}, 8284--8291\relax
\mciteBstWouldAddEndPuncttrue
\mciteSetBstMidEndSepPunct{\mcitedefaultmidpunct}
{\mcitedefaultendpunct}{\mcitedefaultseppunct}\relax
\EndOfBibitem
\bibitem[Andersen \latin{et~al.}(2021)Andersen, Scuri, Sushko, De~Greve, Sung,
  Zhou, Wild, Gelly, Heo, B{\'e}rub{\'e}, \latin{et~al.}
  others]{kim2021excitons}
Andersen,~T.~I.; Scuri,~G.; Sushko,~A.; De~Greve,~K.; Sung,~J.; Zhou,~Y.;
  Wild,~D.~S.; Gelly,~R.~J.; Heo,~H.; B{\'e}rub{\'e},~D., \latin{et~al.}
  Excitons in a reconstructed moir{\'e} potential in twisted WSe2/WSe2
  homobilayers. \emph{Nature Materials} \textbf{2021}, \emph{20},
  480--487\relax
\mciteBstWouldAddEndPuncttrue
\mciteSetBstMidEndSepPunct{\mcitedefaultmidpunct}
{\mcitedefaultendpunct}{\mcitedefaultseppunct}\relax
\EndOfBibitem
\bibitem[Wang \latin{et~al.}(2021)Wang, Zhu, Seyler, Rivera, Zheng, Wang, He,
  Taniguchi, Watanabe, Yan, \latin{et~al.} others]{xiaodong2021moire}
Wang,~X.; Zhu,~J.; Seyler,~K.~L.; Rivera,~P.; Zheng,~H.; Wang,~Y.; He,~M.;
  Taniguchi,~T.; Watanabe,~K.; Yan,~J., \latin{et~al.}  Moir{\'e} trions in
  MoSe2/WSe2 heterobilayers. \emph{Nature Nanotechnology} \textbf{2021},
  \emph{16}, 1208--1213\relax
\mciteBstWouldAddEndPuncttrue
\mciteSetBstMidEndSepPunct{\mcitedefaultmidpunct}
{\mcitedefaultendpunct}{\mcitedefaultseppunct}\relax
\EndOfBibitem
\bibitem[Shabani \latin{et~al.}(2021)Shabani, Halbertal, Wu, Chen, Liu, Hone,
  Yao, Basov, Zhu, and Pasupathy]{shabani2021deep}
Shabani,~S.; Halbertal,~D.; Wu,~W.; Chen,~M.; Liu,~S.; Hone,~J.; Yao,~W.;
  Basov,~D.~N.; Zhu,~X.; Pasupathy,~A.~N. Deep moir{\'e} potentials in twisted
  transition metal dichalcogenide bilayers. \emph{Nature Physics}
  \textbf{2021}, \emph{17}, 720--725\relax
\mciteBstWouldAddEndPuncttrue
\mciteSetBstMidEndSepPunct{\mcitedefaultmidpunct}
{\mcitedefaultendpunct}{\mcitedefaultseppunct}\relax
\EndOfBibitem
\bibitem[Bai \latin{et~al.}(2020)Bai, Zhou, Wang, Wu, McGilly, Halbertal, Lo,
  Liu, Ardelean, Rivera, \latin{et~al.} others]{XYZ:2019:1d}
Bai,~Y.; Zhou,~L.; Wang,~J.; Wu,~W.; McGilly,~L.~J.; Halbertal,~D.; Lo,~C.
  F.~B.; Liu,~F.; Ardelean,~J.; Rivera,~P., \latin{et~al.}  Excitons in
  strain-induced one-dimensional moir{\'e} potentials at transition metal
  dichalcogenide heterojunctions. \emph{Nature Materials} \textbf{2020},
  \emph{19}, 1068--1073\relax
\mciteBstWouldAddEndPuncttrue
\mciteSetBstMidEndSepPunct{\mcitedefaultmidpunct}
{\mcitedefaultendpunct}{\mcitedefaultseppunct}\relax
\EndOfBibitem
\bibitem[Zhang \latin{et~al.}(2021)Zhang, Regan, Wang, Zhao, Wang, Sayyad,
  Yumigeta, Watanabe, Taniguchi, Tongay, \latin{et~al.}
  others]{crommie2021correlated}
Zhang,~Z.; Regan,~E.~C.; Wang,~D.; Zhao,~W.; Wang,~S.; Sayyad,~M.;
  Yumigeta,~K.; Watanabe,~K.; Taniguchi,~T.; Tongay,~S., \latin{et~al.}
  Correlated interlayer exciton insulator in double layers of monolayer WSe2
  and moir$\backslash$'e WS2/WSe2. \emph{arXiv preprint arXiv:2108.07131}
  \textbf{2021}, \relax
\mciteBstWouldAddEndPunctfalse
\mciteSetBstMidEndSepPunct{\mcitedefaultmidpunct}
{}{\mcitedefaultseppunct}\relax
\EndOfBibitem
\bibitem[Darlington \latin{et~al.}(2020)Darlington, Carmesin, Florian, Yanev,
  Ajayi, Ardelean, Rhodes, Ghiotto, Krayev, Watanabe, \latin{et~al.}
  others]{darlington2020imaging}
Darlington,~T.~P.; Carmesin,~C.; Florian,~M.; Yanev,~E.; Ajayi,~O.;
  Ardelean,~J.; Rhodes,~D.~A.; Ghiotto,~A.; Krayev,~A.; Watanabe,~K.,
  \latin{et~al.}  Imaging strain-localized excitons in nanoscale bubbles of
  monolayer WSe2 at room temperature. \emph{Nature Nanotechnology}
  \textbf{2020}, \emph{15}, 854--860\relax
\mciteBstWouldAddEndPuncttrue
\mciteSetBstMidEndSepPunct{\mcitedefaultmidpunct}
{\mcitedefaultendpunct}{\mcitedefaultseppunct}\relax
\EndOfBibitem
\bibitem[Carmesin \latin{et~al.}(2019)Carmesin, Lorke, Florian, Erben, Schulz,
  Wehling, and Jahnke]{carmesin2019quantum}
Carmesin,~C.; Lorke,~M.; Florian,~M.; Erben,~D.; Schulz,~A.; Wehling,~T.~O.;
  Jahnke,~F. Quantum-dot-like states in molybdenum disulfide nanostructures due
  to the interplay of local surface wrinkling, strain, and dielectric
  confinement. \emph{Nano letters} \textbf{2019}, \emph{19}, 3182--3186\relax
\mciteBstWouldAddEndPuncttrue
\mciteSetBstMidEndSepPunct{\mcitedefaultmidpunct}
{\mcitedefaultendpunct}{\mcitedefaultseppunct}\relax
\EndOfBibitem
\bibitem[Chirolli \latin{et~al.}(2019)Chirolli, Prada, Guinea, Rold{\'a}n, and
  San-Jose]{chirolli2019strain}
Chirolli,~L.; Prada,~E.; Guinea,~F.; Rold{\'a}n,~R.; San-Jose,~P.
  Strain-induced bound states in transition-metal dichalcogenide bubbles.
  \emph{2D Materials} \textbf{2019}, \emph{6}, 025010\relax
\mciteBstWouldAddEndPuncttrue
\mciteSetBstMidEndSepPunct{\mcitedefaultmidpunct}
{\mcitedefaultendpunct}{\mcitedefaultseppunct}\relax
\EndOfBibitem
\bibitem[Morrow and Ma(2021)Morrow, and Ma]{morrow2021trapping}
Morrow,~D.~J.; Ma,~X. Trapping interlayer excitons in van der Waals
  heterostructures by potential arrays. \emph{Phys. Rev. B} \textbf{2021},
  \emph{104}, 195302--195312\relax
\mciteBstWouldAddEndPuncttrue
\mciteSetBstMidEndSepPunct{\mcitedefaultmidpunct}
{\mcitedefaultendpunct}{\mcitedefaultseppunct}\relax
\EndOfBibitem
\bibitem[El-Khoury and Schultz(2020)El-Khoury, and Schultz]{Khoury2020acs}
El-Khoury,~P.~Z.; Schultz,~Z.~D. From SERS to TERS and Beyond: Molecules as
  Probes of Nanoscopic Optical Fields. \emph{The Journal of Physical Chemistry
  C} \textbf{2020}, \emph{124}, 27267--27275\relax
\mciteBstWouldAddEndPuncttrue
\mciteSetBstMidEndSepPunct{\mcitedefaultmidpunct}
{\mcitedefaultendpunct}{\mcitedefaultseppunct}\relax
\EndOfBibitem
\bibitem[Tyurnina \latin{et~al.}(2019)Tyurnina, Bandurin, Khestanova, Kravets,
  Koperski, Guinea, Grigorenko, Geim, and Grigorieva]{tyurnina2019strained}
Tyurnina,~A.~V.; Bandurin,~D.~A.; Khestanova,~E.; Kravets,~V.~G.; Koperski,~M.;
  Guinea,~F.; Grigorenko,~A.~N.; Geim,~A.~K.; Grigorieva,~I.~V. Strained
  bubbles in van der Waals heterostructures as local emitters of
  photoluminescence with adjustable wavelength. \emph{ACS photonics}
  \textbf{2019}, \emph{6}, 516--524\relax
\mciteBstWouldAddEndPuncttrue
\mciteSetBstMidEndSepPunct{\mcitedefaultmidpunct}
{\mcitedefaultendpunct}{\mcitedefaultseppunct}\relax
\EndOfBibitem
\bibitem[Sigl \latin{et~al.}(2022)Sigl, Troue, Katzer, Selig, Sigger, Kiemle,
  Brotons-Gisbert, Watanabe, Taniguchi, Gerardot, \latin{et~al.}
  others]{sigl2022optical}
Sigl,~L.; Troue,~M.; Katzer,~M.; Selig,~M.; Sigger,~F.; Kiemle,~J.;
  Brotons-Gisbert,~M.; Watanabe,~K.; Taniguchi,~T.; Gerardot,~B.~D.,
  \latin{et~al.}  Optical dipole orientation of interlayer excitons in MoSe 2-
  WSe 2 heterostacks. \emph{Physical Review B} \textbf{2022}, \emph{105},
  035417\relax
\mciteBstWouldAddEndPuncttrue
\mciteSetBstMidEndSepPunct{\mcitedefaultmidpunct}
{\mcitedefaultendpunct}{\mcitedefaultseppunct}\relax
\EndOfBibitem
\bibitem[Su \latin{et~al.}(2022)Su, Xu, Cheng, Li, Liu, Watanabe, Taniguchi,
  Berkelbach, Hone, and Delor]{su2022dark}
Su,~H.; Xu,~D.; Cheng,~S.-W.; Li,~B.; Liu,~S.; Watanabe,~K.; Taniguchi,~T.;
  Berkelbach,~T.~C.; Hone,~J.~C.; Delor,~M. Dark-exciton driven energy
  funneling into dielectric inhomogeneities in two-dimensional semiconductors.
  \emph{Nano Letters} \textbf{2022}, \emph{22}, 2843--2850\relax
\mciteBstWouldAddEndPuncttrue
\mciteSetBstMidEndSepPunct{\mcitedefaultmidpunct}
{\mcitedefaultendpunct}{\mcitedefaultseppunct}\relax
\EndOfBibitem
\bibitem[Harats \latin{et~al.}(2020)Harats, Kirchhof, Qiao, Greben, and
  Bolotin]{harats2020dynamics}
Harats,~M.~G.; Kirchhof,~J.~N.; Qiao,~M.; Greben,~K.; Bolotin,~K.~I. Dynamics
  and efficient conversion of excitons to trions in non-uniformly strained
  monolayer WS2. \emph{Nature Photonics} \textbf{2020}, \emph{14},
  324--329\relax
\mciteBstWouldAddEndPuncttrue
\mciteSetBstMidEndSepPunct{\mcitedefaultmidpunct}
{\mcitedefaultendpunct}{\mcitedefaultseppunct}\relax
\EndOfBibitem
\bibitem[Lee \latin{et~al.}(2022)Lee, Koo, Choi, Kumar, Lee, Ji, Choi, Kang,
  Kim, Park, \latin{et~al.} others]{lee2022drift}
Lee,~H.; Koo,~Y.; Choi,~J.; Kumar,~S.; Lee,~H.-T.; Ji,~G.; Choi,~S.~H.;
  Kang,~M.; Kim,~K.~K.; Park,~H.-R., \latin{et~al.}  Drift-dominant exciton
  funneling and trion conversion in 2D semiconductors on the nanogap.
  \emph{Science advances} \textbf{2022}, \emph{8}, 5236\relax
\mciteBstWouldAddEndPuncttrue
\mciteSetBstMidEndSepPunct{\mcitedefaultmidpunct}
{\mcitedefaultendpunct}{\mcitedefaultseppunct}\relax
\EndOfBibitem
\bibitem[Kamban and Pedersen(2020)Kamban, and Pedersen]{kamban2020interlayer}
Kamban,~H.~C.; Pedersen,~T.~G. Interlayer excitons in van der Waals
  heterostructures: Binding energy, Stark shift, and field-induced
  dissociation. \emph{Scientific reports} \textbf{2020}, \emph{10}, 1--10\relax
\mciteBstWouldAddEndPuncttrue
\mciteSetBstMidEndSepPunct{\mcitedefaultmidpunct}
{\mcitedefaultendpunct}{\mcitedefaultseppunct}\relax
\EndOfBibitem
\bibitem[Bussy \latin{et~al.}(2017)Bussy, Pizzi, and
  Gibertini]{bussy2017strain}
Bussy,~A.; Pizzi,~G.; Gibertini,~M. Strain-induced polar discontinuities in
  two-dimensional materials from combined first-principles and
  Schr{\"o}dinger-Poisson simulations. \emph{Physical Review B} \textbf{2017},
  \emph{96}, 165438\relax
\mciteBstWouldAddEndPuncttrue
\mciteSetBstMidEndSepPunct{\mcitedefaultmidpunct}
{\mcitedefaultendpunct}{\mcitedefaultseppunct}\relax
\EndOfBibitem
\bibitem[Korm{\'a}nyos \latin{et~al.}(2015)Korm{\'a}nyos, Burkard, Gmitra,
  Fabian, Z{\'o}lyomi, Drummond, and Fal’ko]{kormanyos2015k}
Korm{\'a}nyos,~A.; Burkard,~G.; Gmitra,~M.; Fabian,~J.; Z{\'o}lyomi,~V.;
  Drummond,~N.~D.; Fal’ko,~V. k{\textperiodcentered} p theory for
  two-dimensional transition metal dichalcogenide semiconductors. \emph{2D
  Materials} \textbf{2015}, \emph{2}, 022001\relax
\mciteBstWouldAddEndPuncttrue
\mciteSetBstMidEndSepPunct{\mcitedefaultmidpunct}
{\mcitedefaultendpunct}{\mcitedefaultseppunct}\relax
\EndOfBibitem
\bibitem[Duerloo \latin{et~al.}(2012)Duerloo, Ong, and
  Reed]{duerloo2012intrinsic}
Duerloo,~K.-A.~N.; Ong,~M.~T.; Reed,~E.~J. Intrinsic piezoelectricity in
  two-dimensional materials. \emph{The Journal of Physical Chemistry Letters}
  \textbf{2012}, \emph{3}, 2871--2876\relax
\mciteBstWouldAddEndPuncttrue
\mciteSetBstMidEndSepPunct{\mcitedefaultmidpunct}
{\mcitedefaultendpunct}{\mcitedefaultseppunct}\relax
\EndOfBibitem
\bibitem[Li \latin{et~al.}(2022)Li, Xiao, Guan, Xiao, Xu, Zhang, Lin, Li, Tong,
  Li, and Pan]{Listrain}
Li,~K.; Xiao,~F.; Guan,~W.; Xiao,~Y.; Xu,~C.; Zhang,~J.; Lin,~C.; Li,~D.;
  Tong,~Q.; Li,~S.-Y.; Pan,~A. Morphology Deformation and Giant Electronic Band
  Modulation in Long-Wavelength WS2 Moiré Superlattices. \emph{Nano Letters}
  \textbf{2022}, \emph{22}, 5997--6003, PMID: 35839083\relax
\mciteBstWouldAddEndPuncttrue
\mciteSetBstMidEndSepPunct{\mcitedefaultmidpunct}
{\mcitedefaultendpunct}{\mcitedefaultseppunct}\relax
\EndOfBibitem
\bibitem[Rizzo \latin{et~al.}(2022)Rizzo, Shabani, Jessen, Zhang, McLeod,
  Rubio-Verdú, Ruta, Cothrine, Yan, Mandrus, Nagler, Rubio, Hone, Dean,
  Pasupathy, and Basov]{rucl3}
Rizzo,~D.~J. \latin{et~al.}  Nanometer-Scale Lateral p–n Junctions in
  Graphene/$\alpha$-RuCl3 Heterostructures. \emph{Nano Letters} \textbf{2022},
  \emph{22}, 1946--1953, PMID: 35226804\relax
\mciteBstWouldAddEndPuncttrue
\mciteSetBstMidEndSepPunct{\mcitedefaultmidpunct}
{\mcitedefaultendpunct}{\mcitedefaultseppunct}\relax
\EndOfBibitem
\bibitem[Balgley \latin{et~al.}(2022)Balgley, Butler, Biswas, Ge, Lagasse,
  Taniguchi, Watanabe, Cothrine, Mandrus, Velasco~Jr, \latin{et~al.}
  others]{balgley2022ultra}
Balgley,~J.; Butler,~J.; Biswas,~S.; Ge,~Z.; Lagasse,~S.; Taniguchi,~T.;
  Watanabe,~K.; Cothrine,~M.; Mandrus,~D.~G.; Velasco~Jr,~J., \latin{et~al.}
  Ultra-sharp lateral pn junctions in modulation-doped graphene. \emph{arXiv
  preprint arXiv:2203.06295} \textbf{2022}, \relax
\mciteBstWouldAddEndPunctfalse
\mciteSetBstMidEndSepPunct{\mcitedefaultmidpunct}
{}{\mcitedefaultseppunct}\relax
\EndOfBibitem
\end{mcitethebibliography}
\section{Data availability}
The data used in this paper are available from the corresponding authors upon reasonable request.

\section{Acknowledgement}
Single crystal synthesis and heterostructure fabrication for this work was supported by the the NSF MRSEC program through Columbia in the Center for Precision-Assembled Quantum Materials (PAQM), grant number DMR-2011738. STM measurements were supported by the NSF GOALI program under grant DMR-1809122 and by the Air Force Office of ScientificResearch via grant FA9550-16-1-0601. CED and CG acknowledge support from the National Science Foundation under Grant No. DMR-1918455. The Flatiron Institute is a division of the Simons Foundation. PJS, TD, and EY acknowledge support from the National Science Foundation through award DMR-2004437.

\section{Author Contributions}
S.S. carried out the STM/STS measurements and analysis.
T.D. conducted optical measurements. T.D. and S.S. analyzed the nano-optical data. W.W. and E.Y fabricated the device and performed SHG experiments. C.G. performed the theoretical calculations. J.H., X.Z., C.E.D., P.J.S. and A.N.P. provided advice. All authors contributed to writing the manuscript.

\section{Competing interests}
The authors declare no competing interests.


\newpage
\begin{figure*}
\centering
\includegraphics[width=1\linewidth]{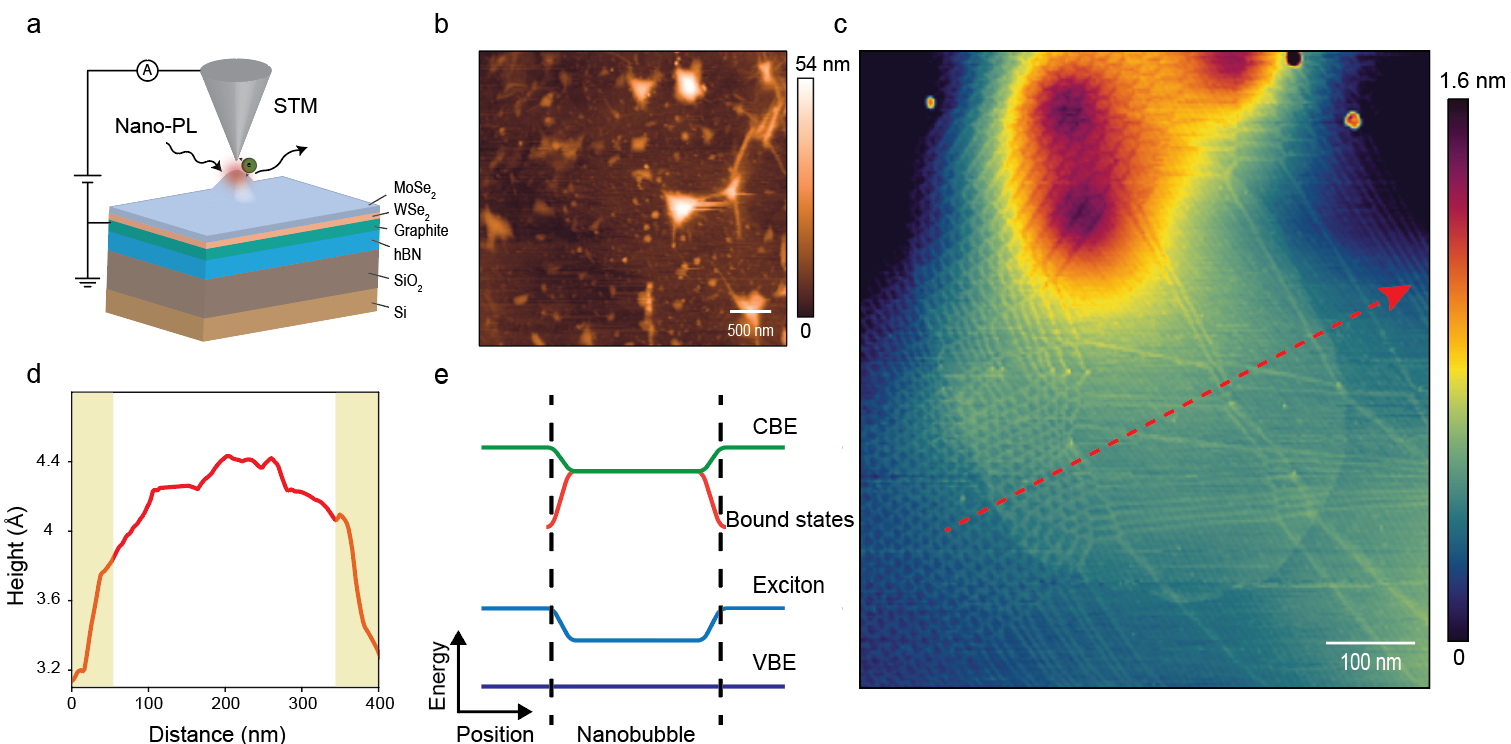}                  \vspace{-0.15 in}
\caption{\small{{\bf{Nanobubbles in twisted \wse2/\mose2 heterobilayers}}
{\bf{a}} Schematic of the experimental setup.
{\bf{b}} 3.8 micron AFM image of a \wse2/\mose2 heterobilayer on gold substrate used in nano-PL measurements showing several nanobubbles {\bf{c}} STM topographic image ($V_b$ = -1.8 V, I = -50 pA), of a single nanobubble. The continuous \moire pattern across the nanobubble shows that contact between the two TMD layers is maintained throughout. {\bf{d}} STM Height profile across the dashed arrow in (c) showing an apparent step size of 1 angstrom at the nanobubble edge highlighted in yellow.
{\bf{e}} Schematic band diagram of the conduction band edge (CB) , the bound states, the valence band edge(VB), and interlayer exciton energy as a function of position deduced from STM and nano-PL measurements.}}                                                     

\label{fig1}
\vspace{-0.15 in}
\end{figure*}
\newpage

\begin{figure*}
\centering
\includegraphics[width=1\linewidth]{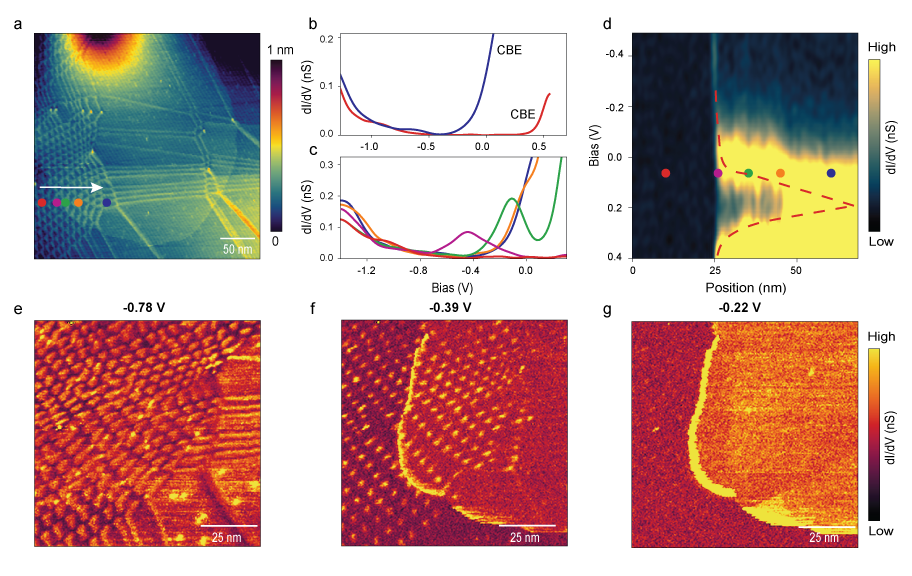}                                                                                                                                                
\vspace{-0.15 in}

\caption{\small{
{\bf{STM spectroscopic properties of nanobubbles}}
{\bf{a}} STM topography indicating locations where the point spectra shown in {\bf{b,c}} were taken. {\bf{b}} Point spectra taken inside (blue) and outside (Red) the bubble show a pronounced shift of the conduction band edge (CBE) inside the bubble while the valence band is unaffected.
{\bf{c}} Evolution of \didv point spectra from the outside to the inside of the nanobubble corresponding to the markers in {\bf{a}}. At the edge of the bubble, deeply localized states are seen within the semiconductor gap.  {\bf{d}} Heat map of position dependent \didv  spectra at the conduction band edge across the bubble. The dashed lines are guides to the eye that show the evolution of the bound states and the conduction band edge. {\bf{e-g}} Spectroscopic maps across the nanobubble edge at energies of -0.78 V (valence band edge), -0.39 V, and -0.22 V (within the bandgap) The set points are $V_b$ = -1.8 V, I = -200 pA.  Bound states are clearly observed at the nanobubble edge at -0.39 V and -0.22 V.
}}   
\label{fig1}
\vspace{-0.15 in}
\end{figure*}
\newpage

\begin{figure*}
\centering
\includegraphics[width=1\linewidth]{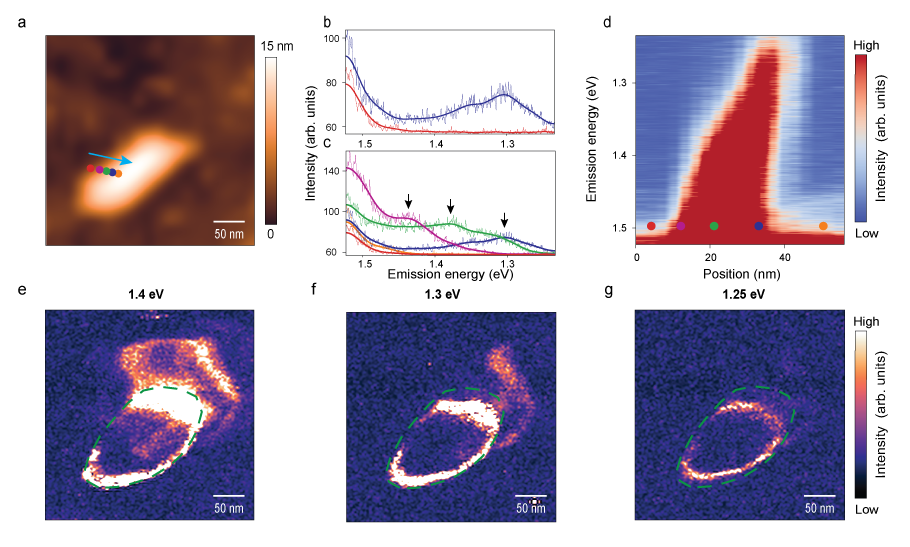}                                                                                                                                                
\vspace{-0.15 in}
\caption{\small{
{\bf{Hyperspectral nano-PL maps of localized interlayer exciton spectra}}
{\bf{a}} Atomic force micrograph of \mose2-\wse2 nanobubbles on a template-stripped Au
substrate. {\bf{b}} Average nano-PL spectra on (blue) and off (red) the nanobubble region.
{\bf{c}} Nano-PL point spectra across the bubble edge corresponding the colored points in {\bf{a}}. For points farther into the interior of the
nanobubble, the emissions energy redshifts.
{\bf{d}} Hyperspectral heat map along the vector defined by points in (a). The redshifting of the emission is
clearly seen, with the spectral weight shifting from above 1.5 eV at the very edge of the bubble (red point) to below 1.3 eV at 35 nm inside the bubble (blue point). Past the position past 40 nm (orange point) the intensity shift
below the low energy detection cutoff of the EMCCD of ~1.2 eV.
{\bf{e-g}} nano-PL maps (average with a window sized 2.5 nm) binned different energy intervals, ordered by decreasing emission energy. The approximate physical nanobubble edge is shown with the green dashed line. The data clearly shows the redshifting of the emission on going into the bubble.}}   
\label{fig1}
\vspace{-0.15 in}
\end{figure*}
\newpage

\begin{figure*}
\centering
\includegraphics[width=1\linewidth]{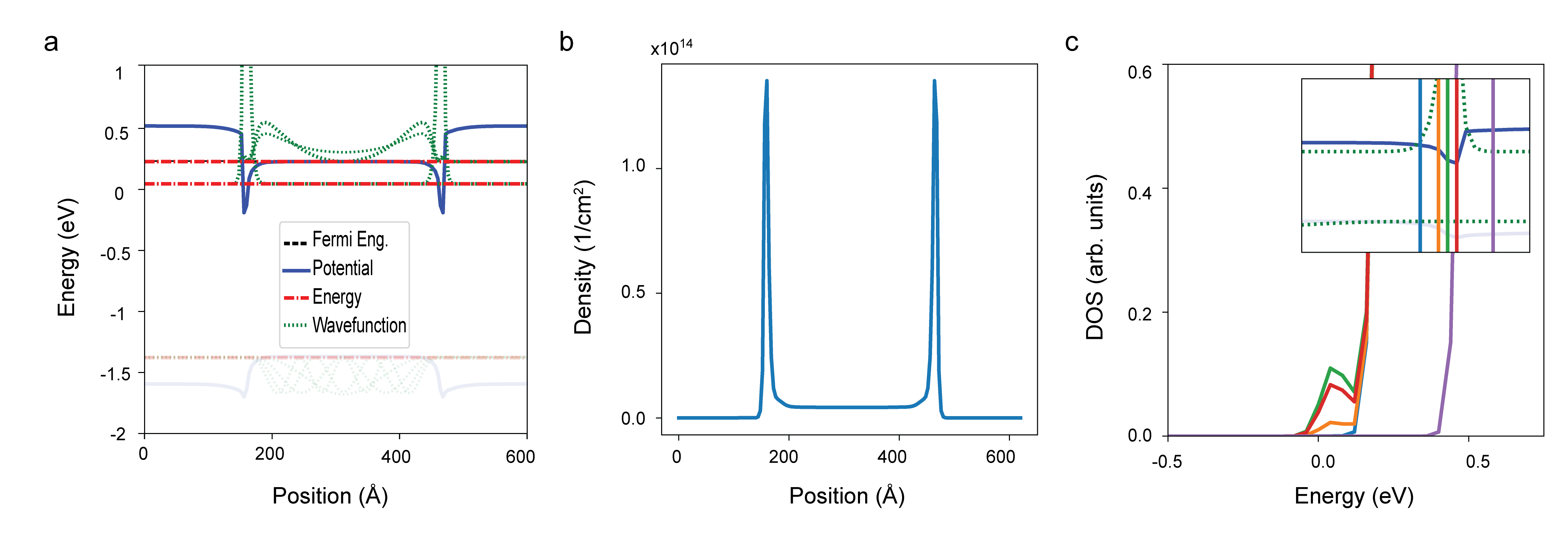}                                                                                                                                                
\vspace{-0.15 in}
\caption{\small{
{\bf{Schrödinger-Poisson (SP) simulations}}
{\bf{a}} Band diagram for two-dimensional Schrödinger-Poisson (SP) model of the bubble. Blue solid
lines are the self-consistent band potentials, red dot-dashed lines are the energies of some of the confined
states, black dashed line is the Fermi level, and green dotted lines are the confined
wavefunctions. {\bf{b}} Self-consistent electron density versus $x$ position in (a). {\bf{c}} Density of states versus energy including the 1D states confined to the well, with 2D DOS outside of the bubble, and in the
bubble interior, superimposed. }}   
\label{fig1}
\vspace{-0.15 in}
\end{figure*}

\end{document}